\documentclass[pra,twocolumn,showpacs,amsmath,amssymb,nofootinbib]{revtex4}
\usepackage{color}
\usepackage{graphicx}
\usepackage{dcolumn}
\usepackage{bm}

\newcommand{\vek}[1]{\bm{\mathrm{#1}}}

\newcommand{\kv}{\vek{k}}
\newcommand{\pv}{\vek{p}}

\newcommand{\qv}{\vek{q}}
\newcommand{\rv}{\vek{r}}

\newcommand{\eq}{\textit{eq}}
\newcommand{\cl}{\textit{cl}}
\newcommand{\ho}{\textit{ho}}
\newcommand{\coll}{\textit{coll}}
\newcommand{\trans}{\textit{trans}}
\newcommand{\hd}{\textit{hd}}
\newcommand{\Eq}[1]{Eq.\@ (\ref{#1})}
\newcommand{\Eqs}[1]{Eqs.\@ (\ref{#1})}
\newcommand{\Ref}[1]{Ref.\@ \cite{#1}}
\newcommand{\Refs}[1]{Refs.\@ \cite{#1}}
\newcommand{\Fig}[1]{Fig.\@ \ref{#1}}

\newcommand{\Sec}[1]{Sec.\@ \ref{#1}}
\newcommand{\Tab}[1]{Table\@ \ref{#1}}

\DeclareMathOperator{\Imag}{Im}

\renewcommand{\Im}{\Imag}
\begin{document}

\title{Role of fourth-order phase-space moments in collective modes of
  trapped Fermi gases}

\author{Silvia Chiacchiera}
\affiliation{Centro de F{\'i}sica Computacional, Department of Physics,
University of Coimbra, P-3004-516 Coimbra, Portugal}
\author{Thomas Lepers}
\affiliation{Universit{\'e} de Lyon, F-69622 Lyon, France;
  Univ. Lyon 1, Villeurbanne;
  CNRS/IN2P3, UMR5822, IPNL}
\author{Dany Davesne}
\affiliation{Universit{\'e} de Lyon, F-69622 Lyon, France;
  Univ. Lyon 1, Villeurbanne;
  CNRS/IN2P3, UMR5822, IPNL}
\author{Michael Urban}
\affiliation{Institut de Physique Nucl{\'e}aire, CNRS/IN2P3 and
  Univ. Paris-Sud 11, 91406 Orsay cedex, France}
\date{July 5, 2011}
\begin{abstract}
We study the transition from hydrodynamic to collisionless behavior in
collective modes of ultracold trapped Fermi gases. To that end, we
solve the Boltzmann equation for the trapped Fermi gas via the moments
method. We showed previously that it is necessary to go beyond
second-order moments if one wants to reproduce the results of a
numerical solution of the Boltzmann equation. Here, we will give the
detailed description of the method including fourth-order moments. We
apply this method to the case of realistic parameters, and compare the
results for the radial quadrupole and scissors modes at unitarity to
experimental data obtained by the Innsbruck group. It turns out that
the inclusion of fourth-order moments clearly improves the agreement
with the experimental data. In particular, the fourth-order moments
reduce the effect of collisions and therefore partially compensate the
effect of the enhanced in-medium cross section at low temperatures.
\end{abstract}
\pacs{67.85.Lm,03.75.Ss}
\maketitle

\section{\label{sec:intro}Introduction}
The study of collective modes of trapped two-component Fermi gases
revealed interesting information about different dynamical regimes
\cite{Wright}. Initially, the aim was to find signals for the
superfluid-normal phase transition. However, near a Feshbach
resonance, the atom-atom scattering cross section can be large enough
to ensure (normal-fluid) hydrodynamic behavior of the gas above the
superfluid critical temperature $T_c$. In this case, a change in the
behavior of the gas is observed at much higher temperature, when the
gas gets more and more dilute until the collisionless regime is
reached. The most interesting modes in this context are those which in
the collisionless case exhibit deformations in both coordinate and
momentum space. Such modes are, e.g., the quadrupole and the scissors
modes. In the presence of superfluidity or collisions, the deformation
of the momentum sphere is suppressed, so that the frequencies of these
modes are different from those in the collisionless case. In the
intermediate regime, the damping of these modes is very strong. Both
the radial quadrupole mode and the scissors mode were experimentally
studied by the Innsbruck group \cite{Altmeyer,Wright,RiedlBruun}.

From the theoretical side, the continuous transition from
collisionally hydrodynamic to collisionless behavior can be studied by
using the semiclassical Boltzmann equation. At present, there is no
technique which would allow for a fully quantum mechanical description
of collective modes of systems containing a few hundred thousand
particles, including the collisional effects. But even the solution of
the Boltzmann equation is far from being simple, and most of the time
further approximations are made. A very common approximation is the
relaxation-time approximation, which was used, together with the
so-called scaling ansatz, to describe collective modes and the
expansion of the gas after the trap is switched off \cite{Pedri}. In
the case of collective modes, this method is equivalent to the method
of phase-space moments up to second order, which was applied to the
radial quadrupole, scissors, and breathing modes
\cite{BruunSmith07,RiedlBruun,Chiacchiera}. In both methods, the
phase-space distribution function is constrained to a simple form, but
the advantage is that one can perform computations almost
analytically.

There are other approaches like the fully numerical solution of the
Boltzmann equation as developed, e.g., in
\Refs{ToschiVignolo,ToschiCapuzzi,Lepers,Goulko}. In this case no
constraint is put on the functional form of the distribution function,
but the price to pay is that the computations are very time consuming.
Maybe the computation time could be significantly reduced by using new
adaptive algorithms \cite{Wade}, but to our knowledge, no numerical
calculation has been performed so far for degenerate Fermi gases with
parameters (number of atoms, trap geometry) corresponding to real
experiments.

In our previous work \cite{Lepers}, we compared the results of a
numerical simulation of the quadrupole mode in a spherical trap
containing a reduced number of atoms with the corresponding results of
the second-order moments method. The surprising outcome was that the
second-order moments method strongly overestimates the effects of
collisions. This problem could be cured to a large extent by
generalizing the method of moments to fourth order. This is our main
motivation for the present work, where we apply the fourth-order
moments method to the radial quadrupole and scissors modes in a
realistic trap geometry, and compare it directly with the experimental
results of \Ref{RiedlBruun}.

The Boltzmann equation requires as microscopic input the mean-field
potential (in the Vlasov part) and the cross section (in the collision
integral). Here, the mean field will be neglected since we found in
\Ref{Chiacchiera} that it affects only very weakly the frequencies and
damping rates of the collective modes near unitarity. The main effect
comes clearly from the collisions. In the case of large scattering
length $a$ and temperature slightly above $T_c$, one expects the Fermi
gas to be in the ``pseudogap regime'' in which pair correlations play
an important role although the pairs are not condensed. In this
regime, the relaxation time is strongly reduced since the scattering
cross section calculated in the surrounding medium is enhanced as
compared with the free one \cite{BruunSmith05} -- an effect which in
the context of nuclear physics has already been known for a couple of 
years \cite{Alm}. Previous studies \cite{RiedlBruun,Chiacchiera} using
the in-medium cross section in trapped Fermi gases found that this
reduction of the relaxation time badly deteriorates the agreement with
the experimental results. Here we argue that this discrepancy was, at
least to some extent, due to the failure of the second-order moments
method and not due to the enhancement of the in-medium cross section.

The paper is organized as follows. In \Sec{sec:formalism} we describe
the method, starting with a very general formulation and specializing
then to the scissors and radial quadrupole modes. We explain how the
response of the system is obtained and how we extract from it the
frequencies and damping rates. In \Sec{sec:results} we discuss our
results. We show how the inclusion of fourth-order moments affects the
response function and the corresponding frequencies and damping rates
and compare the theoretical results with experimental data. In
\Sec{sec:summary} we summarize and give an outlook to future
studies. Some technical details are given in the appendix.

Throughout the article, we use units with $\hbar = k_B = 1$.

\section{Formalism}
\label{sec:formalism}
\subsection{Moments method for the Boltzmann equation}
We consider a two-component ($\uparrow,\downarrow$) gas of Fermi atoms
of mass $m$ and with an interspecies attractive interaction (the
scattering length is $a<0$). The gas is loaded in a harmonic, usually
anisotropic, trap
\begin{equation}
V(\rv)=\frac{m}{2}(\omega_x^2 x^2 + \omega_y^2 y^2 +\omega_z^2 z^2)~.
\end{equation} 
Moreover, since we are interested in collective modes and their time
evolution, we include in the external potential felt by the atoms a
small time-dependent part $\delta V(\rv,t)$, that will be used to
simulate the excitation of the mode. As mentioned before, the mean
field felt by the atoms due to their interaction will be neglected
here, since at unitarity 
it is only of minor importance for the properties of
collective modes as compared with the effects coming from collisions
between atoms \cite{Chiacchiera}.

In the normal fluid phase and under other assumptions we already
discussed in \Ref{Chiacchiera}, we can describe the system with a
semiclassical distribution function $f_\sigma(\rv,\pv,t)$, where
$\sigma=\uparrow,\downarrow$. We restrict ourselves to the case of an
unpolarized gas ($N_\uparrow=N_\downarrow\equiv N/2$) and to
excitations where the two components move together:
$f_\uparrow=f_\downarrow\equiv f$. The normalization of $f$
is\footnote{Notice that this normalization differs from that given in
  \Ref{Chiacchiera} by a factor $(2\pi)^3$.}
\begin{equation}
  \int \frac{d^3r d^3 p}{(2\pi)^3} f(\rv,\pv,t) = \frac{N}{2}
\end{equation}
and the average value of a generic quantity $\chi(\rv,\pv)$ is
\begin{equation}\label{eq:average}
\langle \chi\rangle=\frac{2}{N}\int \frac{d^3r d^3p}{(2\pi)^3} 
f(\rv,\pv,t) \chi(\rv,\pv)~.
\end{equation}
In equilibrium, the distribution function reads
\begin{equation}
f_{\eq}(\vek{r},\vek{p})=\frac{1}{e^{\beta[p^2/2m+V(\vek{r})-\mu]}+1}\,,
\end{equation}
where $\beta=1/T$ is the inverse of the temperature and $\mu$ is the
chemical potential.

When the system is excited, the time evolution of $f$ is governed by
the Boltzmann equation \cite{Landau10}. We consider small
perturbations $\delta f$ of the distribution function from equilibrium
and write them as
\begin{equation}
\delta f(\vek{r},\vek{p},t) = f_\eq(\vek{r},\vek{p})[1-
  f_\eq(\vek{r},\vek{p})] \Phi(\vek{r},\vek{p},t)\,.
\label{eq:deltaf}
\end{equation}
The function $\Phi(\vek{r},\vek{p},t)$ can be assumed to be smooth
since the fact that $\delta f$ is peaked near the Fermi surface is
already accounted for by the prefactor $f_\eq(1-f_\eq)$. The
linearized Boltzmann equation then reads
\begin{multline}
  f_\eq(1-f_\eq) \Big(\dot{\Phi} +
  \frac{\vek{p}}{m}\cdot\vek{\nabla}_r\Phi -
  \vek{\nabla}_r V\cdot\vek{\nabla}_p\Phi \\
  + \beta\frac{\vek{p}}{m}\cdot\vek{\nabla}_r\delta V
  \Big) = -I[\Phi]\,.
  \label{eq:BoltzLin}
\end{multline}
The linearized collision integral in the right-hand side is
\begin{multline}
  I[\Phi] = \int \frac{d^3 p_1}{(2\pi)^3}\int d\Omega
  \frac{d\sigma}{d\Omega} \frac{|\vek{p}-\vek{p}_1|}{m}
  f_\eq f_{\eq\,1}\\
  \times (1-f_\eq^\prime)(1-f_{\eq\,1}^\prime)
  (\Phi+\Phi_1-\Phi^\prime-\Phi_1^\prime)\,.
  \label{collisionterm}
\end{multline}
The various $f_\eq$ and $\Phi$ are all evaluated at the same $\rv,t$
but at different momenta $\pv$, $\pv_1$, $\pv^\prime$, or
$\pv_1^\prime$, respectively, which due to momentum and energy
conservation satisfy $\pv+\pv_1 = \pv^\prime + \pv_1^\prime \equiv
\kv$ and $|\pv-\pv_1|=|\pv^\prime-\pv_1^\prime|\equiv 2q$. The solid
angle between the initial and final relative momenta in the
center-of-mass frame, $\qv$ and $\qv^\prime$, is denoted $\Omega$. The
cross-section $d\sigma/d\Omega$ used in the present paper is the
in-medium cross section which is calculated as described in
\Ref{Chiacchiera}. At temperatures close to the superfluid transition
temperature $T_c$, this cross-section is strongly enhanced with
respect to the free one $d\sigma_0/d\Omega = a^2/[1+(qa)^2]$, at least
for collision partners near the Fermi surface with zero total
momentum.

Since the function $\Phi$ is supposed to be smooth, one can try to
approximate it by a polynomial in the components of $\rv$ and $\pv$
with time-dependent coefficients $c_i$,
\begin{equation}
\Phi(\vek{r},\vek{p},t)=
\sum_{i=1}^n c_i(t) \phi_i(\vek{r},\vek{p})~.
\label{eq:Phigeneral}\end{equation}
The choice of the basis functions $\phi_i$ depends on the mode
one wants to describe (see discussions in
\Refs{Khawaja,Chiacchiera}). However, let us first explain the general
idea before focusing on the examples of the radial quadrupole and
scissors modes.

In order to obtain the so-called response function, it is sufficient
to consider a perturbation which is a $\delta$ pulse, i.e.,
\begin{equation}
\delta V(\vek{r},t) = \delta(t)\hat{V}(\vek{r})\,.
\end{equation}
Then the Fourier transform of \Eq{eq:BoltzLin} with respect to $t$
gives
\begin{multline}
  \sum_{i=1}^n c_i(\omega)\Big[ f_\eq(1-f_\eq)\Big(-i\omega \phi_i +
    \frac{\vek{p}}{m}\cdot\vek{\nabla}_r \phi_i - \vek{\nabla}_r
    V\cdot\vek{\nabla}_p \phi_i\Big)\\
 + I[\phi_i] \Big] = -f_\eq (1-f_\eq) \beta \frac{\vek{p}}{m} \cdot
   \vek{\nabla}_r \hat{V}(\vek{r})\,,
\label{eq:BoltzLinFT}
\end{multline}
where $c_i(\omega)$ is the Fourier transform of $c_i(t)$. Now we take
the moments of \Eq{eq:BoltzLinFT}, i.e., we multiply it by each of the
basis functions $\phi_i$ and integrate over phase space. In this way,
we obtain $n$ coupled linear algebraic equations for the $n$
coefficients $c_i(\omega)$. In matrix form, they can be written as
\begin{equation}\label{eq:Ac=a}
\sum_{j=1}^n A_{ij} c_j(\omega)= a_i\,,
\end{equation}
where 
\begin{gather}
A_{ij} = -i\omega M_{ij} + A^{\trans}_{ij} +
  A^{\coll}_{ij}\,, \label{eq:Aij}\\
M_{ij} = \int\frac{d^3rd^3p}{(2\pi)^3} f_\eq(1-f_\eq) \phi_i \phi_j\,,
  \label{eq:Mij}\\
A^{\trans}_{ij} = \int \frac{d^3 r d^3 p}{(2\pi)^3} f_\eq(1-f_\eq)
  \phi_i \Big\{\phi_j,\frac{p^2}{2m}+V\Big\}\,,\label{eq:Atransij}\\
A^{\coll}_{ij} = \int \frac{d^3 r d^3 p}{(2\pi)^3}\phi_i
  I[\phi_j]\,,\label{eq:Acollij}
\end{gather}
and
\begin{equation}
a_i = - \frac{\beta}{m} \int\frac{d^3r d^3 p}{(2\pi)^3}\phi_i
f_\eq(1-f_\eq) \vek{p}\cdot\vek{\nabla}_r \hat{V}(\vek{r})\,.\label{eq:ai}
\end{equation}
The contribution $A^{\trans}_{ij}$ of the transport part of the
Boltzmann equation to $A_{ij}$ has been written in a compact form
using the Poisson brackets $\{\cdot,\cdot\}$. One can show that $M$
and $A^{\coll}$ are symmetric matrices, while $A^{\trans}$ is
antisymmetric.

Once we have solved \Eq{eq:Ac=a} for the coefficients $c_i(\omega)$,
we know the time-dependent distribution function $f_\eq+\delta f$ and we
can obtain the time evolution of the average of any dynamical
quantity using \Eq{eq:average}.

In summary, making a polynomial ansatz for the time-dependent
distribution function, we reduced the linearized Boltzmann equation
from an integro-differential equation to a system of $n$ coupled linear
algebraic equations for the coefficients $c_i$.
\subsection{Scissors and quadrupole modes}
Consider an elongated trap with elliptic transverse section (i.e.,
$\omega_x > \omega_y \gg \omega_z$) containing a gas in equilibrium.
The scissors mode is a collective mode that is excited by tilting the
trap by a small angle ($\simeq 5^\circ$) around the $z$-axis. After
this excitation, the cloud is rotating back and forth around the $z$
axis, and what is measured is the time dependence of the angle of the
orientation of the oscillating cloud with respect to the trap
potential. For the details on the experimental realization of this
mode and the results at finite temperature and different scattering
lengths, see \Refs{Wright,RiedlBruun}.

If the initial potential is harmonic, the scissors mode is excited by
the perturbation
\begin{equation}
\hat{V}(\vek{r})= \alpha\, xy\,,
\end{equation}
where $\alpha$ is a factor characterizing the strength of the
perturbation. Under the assumption that the shape of the cloud does
not change during the oscillation, the measured angle is proportional
to the expectation value
\begin{equation}
Q(t) = \langle xy\rangle\,.
\end{equation}

The minimal ansatz for the function $\Phi$ that can reproduce the
scissors mode contains four terms and reads
\cite{BruunSmith07,RiedlBruun,Chiacchiera}
\begin{equation}
\Phi_{2\textit{nd}} = c_1 xy+c_2 p_x p_y+c_3 x p_y + c_4 y p_x\, .
\label{Phi2nd}
\end{equation}
All four terms are of second order in the components of $\vek{r}$ and
$\vek{p}$. In fact, at second order, there are no other combinations
which satisfy the symmetry of this excitation which is odd under
$(x,p_x)\to (-x,-p_x)$, odd under $(y,p_y)\to(-y,-p_y)$, and even
under $(z,p_z)\to(-z,-p_z)$. In the present case of a harmonic
potential without mean field, this set of basis functions is closed
with respect to the operators that are in the transport part of the
Boltzmann equation, i.e., on the left-hand side of \Eq{eq:BoltzLin}.

As noted in \Ref{Lepers} in the case of the quadrupole mode in a
spherical trap, the method of second-order moments strongly
overestimates the collisional effects because it implicitly neglects
the position dependence of the relaxation time $\tau$. Remember that
the effect of collisions is to produce hydrodynamic behavior by
maintaining the momentum distribution spherical. The deformation of
the momentum distribution is described by the second term in
$\Phi_{2\textit{nd}}$, i.e., the term $\propto p_x p_y$. So, the
corresponding coefficient $c_2$ is large in the case of few or no
collisions and small in the case of frequent collisions. In the
trapped system, however, the collision rate is very different
depending on the position: Near the center, the density and thus the
collision rate is much higher than at the surface. Therefore, the
deformation of the momentum distribution should depend on the
position. This cannot be accomplished with the ansatz (\ref{Phi2nd}),
since the term $\propto p_x p_y$ is independent of $\vek{r}$.

Let us therefore go to the next higher order, which is fourth order.
At this order, terms like $x^2 p_x p_y$ etc. appear which allow us to
describe the position dependence of the deformation of the momentum
distribution. Keeping all terms which respect the symmetries mentioned
above, we must then include 32 terms into the ansatz for $\Phi$:
\begin{equation}\label{eq:Phi4th}
\Phi_{4\textit{th}}(\vek{r},\vek{p},t) = \sum_{i=1}^{32}
  c_i(t) \phi_i(\vek{r},\vek{p})\,.
\end{equation}
The basis functions $\phi_i$ can be compactly defined in the following
way:
\begin{equation}\label{eq:phigh}
\phi_{i+4(j-1)}(\vek{r},\vek{p}) = g_i(\vek{r},\vek{p})
  h_j(\vek{r},\vek{p})\,,
\end{equation}
where $i=1,\dots,4$ and $j=1,\dots,8$, and
\begin{gather}\label{eq:gh}
g_1=x y\,,\quad g_2=p_x p_y\,,\quad g_3= x p_y\,, \quad g_4=y p_x\nonumber\\
h_1=1\,,\quad h_2=x^2\,,\quad h_3=y^2\,,\quad h_4=z^2\nonumber\\
h_5=p_x^2\,,\quad h_6=p_y^2\,,\quad h_7=p_z^2\,,\quad h_8=z p_z\,.
\end{gather}
It is easily seen that the first four terms of $\Phi_{4\textit{th}}$
reproduce $\Phi_{2\textit{nd}}$, while the subsequent ones are
fourth-order terms in the components of $\vek{r}$ and $\vek{p}$.

Let us now turn to another mode, the radial quadrupole mode in an
axially symmetric trap, $\omega_x = \omega_y$. In
\Refs{RiedlBruun,Chiacchiera}, the corresponding perturbation was
written as $\hat{V}\propto x^2-y^2$ and the measured observable was
$\langle x^2-y^2\rangle$. However, since the trap is axially
symmetric, we can rotate the coordinate system by $45^\circ$ around
the $z$ axis without changing anything. By doing so, one sees
immediately that the perturbation is then of the form $\hat{V}\propto
xy$ and the measured observable becomes $\langle xy \rangle$, like for
the scissors mode. In conclusion, the radial quadrupole mode is a
special case of the scissors mode in the limit of equal trap
frequencies $\omega_x=\omega_y$, and it therefore does not require any
additional effort to describe both modes.
\subsection{Response function}
As already mentioned, we follow the observable $Q = \langle
xy\rangle$, which, with our choice of basis functions, can be written
as $Q = \langle \phi_1\rangle$. Using \Eqs{eq:deltaf} and
(\ref{eq:Phigeneral}), this expectation value can be expressed in
terms of the coefficients $c_i$ as
\begin{equation}
Q(\omega)=\frac{2}{N}\sum_{i=1}^{32}M_{1i}c_i(\omega)\,,
\label{eq:Qomegageneral}
\end{equation}
where $M_{1i}$ are the elements of the first row of the matrix $M$
defined in \Eq{eq:Mij}.

Also the vector $a$ on the right-hand side of the linear system of
equations (\ref{eq:Ac=a}) for the coefficients $c_i(\omega)$ can be
expressed with the help of the matrix $M$. Note that
$\vek{p}\cdot\vek{\nabla}_r \hat{V}(\vek{r}) = \alpha (xp_y+yp_x) =
\alpha(\phi_3+\phi_4)$, so that \Eq{eq:ai} becomes
\begin{equation}
a_i=-\frac{\alpha\beta}{m}(M_{i3}+M_{i4})\,.\label{eq:aiexplicit}
\end{equation}

Now, the linear system of equations (\ref{eq:Ac=a}) for the
coefficients $c_i$ has to be solved. After some algebra (see
appendix), the result for the response function can be written as
\begin{equation}
Q(\omega) = \frac{-2i\alpha\beta}{Nm}\sum_{k=1}^n \frac{(MP)_{1k}
  [(P^{-1})_{k3}+(P^{-1})_{k4}]}{\omega-\omega_k+i\Gamma_k}\,,
\label{responsefunction}
\end{equation}
where $\Gamma_k+i\omega_k$ is the $k$th eigenvalue of the matrix
$M^{-1}(A^\trans+A^\coll)$ and $P$ is the matrix
containing in its columns the corresponding eigenvectors.

It should be pointed out that it is a very tedious work to calculate
the elements of the matrices $M$, $A^\trans$ and $A^\coll$
corresponding to the fourth-order moments. Here, we made use of the
Mathematica software to derive the expressions. After that, the actual
numerical calculations are quite fast, the only time-consuming part is
the Monte-Carlo integration of the moments of the collision term in
$A^\coll$. The numerical inversion and diagonalization of a $4\times
4$ (second-order method) or $32\times 32$ (fourth-order method) matrix
does not pose any problem. More details about the calculation of the
matrices are given in the appendix.

For the discussion, the imaginary part of $Q(\omega)$ is particularly
useful, since this so-called strength function describes the
excitation spectrum corresponding to the mode under consideration.
\subsection{Frequencies and damping rates}
\label{sec:fit}
In the previous literature \cite{RiedlBruun,Chiacchiera}, where the
second-order moment method was used, the frequencies and damping rates
of the collective modes were identified with the real and imaginary
parts of the solutions of the characteristic equation $\det A = 0$.
These are of course equal to the imaginary and real part of the
eigenvalues of the matrix $M^{-1}(A^\trans+A^\coll)$ mentioned
above. Now, this method is not applicable any more. At fourth order,
there are many eigenvalues, and sometimes they lie close to each other
and have comparable strength in the response function, so that it is
not clear which one should be chosen.

The question arises what is the physical meaning of several poles if
there is in reality only one damped collective mode. In order to get a
better understanding of this question, let us have a look at a simpler
example, namely a zero-sound wave in a uniform system. For this case,
comparisons between the moments method up to very high order and exact
solutions exist in the literature \cite{Providencia,Watabe}. In the
zero-temperature case, it was found \cite{Providencia} that, with
increasing order of the moments method, the distribution of sharp
peaks in the response function (i.e., poles just below the real
$\omega$ axis) converges to the continuous spectrum (i.e., a branch
cut just below the real $\omega$ axis) of the exact solution of the
Vlasov equation. Hence, in order to extract the Landau damping from
the results of the moments method, one has to consider the
distribution of eigenfrequencies rather than look at their imaginary
parts. In the case of finite temperature \cite{Watabe}, the collisions
provide an additional damping mechanism and they lead to complex
eigenfrequencies.

From the preceding discussion it is clear that the frequency and
damping of a mode cannot be obtained from the real and imaginary parts
of the individual eigenfrequencies given by the moments method, but
that one has to consider the total response function. This point of
view is confirmed by the good agreement between the response functions
obtained by the fourth-order moments method and by numerical
simulations in \Ref{Lepers}.

Besides this theoretical question, there is a more practical point one
should consider. The idea is that we want to compare with experimental
data, which were obtained by fitting the observed oscillation of the
cloud with an exponentially damped cosine function. More precisely, in
the case of the quadrupole mode, the observed oscillation is fitted
with a function of the form \cite{Altmeyer}
\begin{equation}
  Q_\textit{fit}(t) = C_1 e^{-\Gamma t} \cos(\omega
  t+\varphi)+ C_2 e^{-\kappa t}\,,
\label{fitq}
\end{equation}
while in the case of the scissors mode, the oscillation is either
fitted with
\begin{equation}
  Q_\textit{fit}^{\textit{low-}T}(t) = C e^{-\Gamma t} \cos(\omega
  t+\varphi)
\label{fits_low}
\end{equation}
at low temperature (hydrodynamic regime), or with
\begin{equation}
  Q_\textit{fit}^{\textit{high-}T}(t) = \sum_{k=1}^2 C_k e^{-\Gamma_k t}
  \cos(\omega_k t+\varphi_k)
\label{fits_high}
\end{equation}
at high temperature (collisionless regime) \cite{Wright}. So, we will
determine the frequency and damping rate corresponding to our response
function $Q(\omega)$ by fitting it with \Eq{fitq} in the case of the
quadrupole mode and with \Eq{fits_low} or (\ref{fits_high}) in the
case of the scissors mode. In the case of a fit with two frequencies,
we concentrate on the mode with the higher frequency.
\section{Results}
\label{sec:results}
\subsection{Scissors and quadrupole strength functions}
\begin{figure*}
\includegraphics[width=5.9cm]{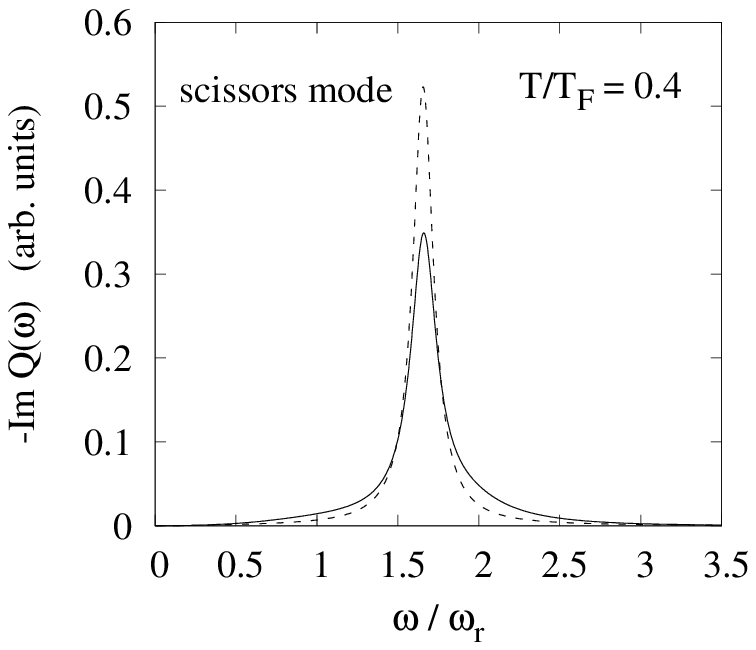} 
\includegraphics[width=5.9cm]{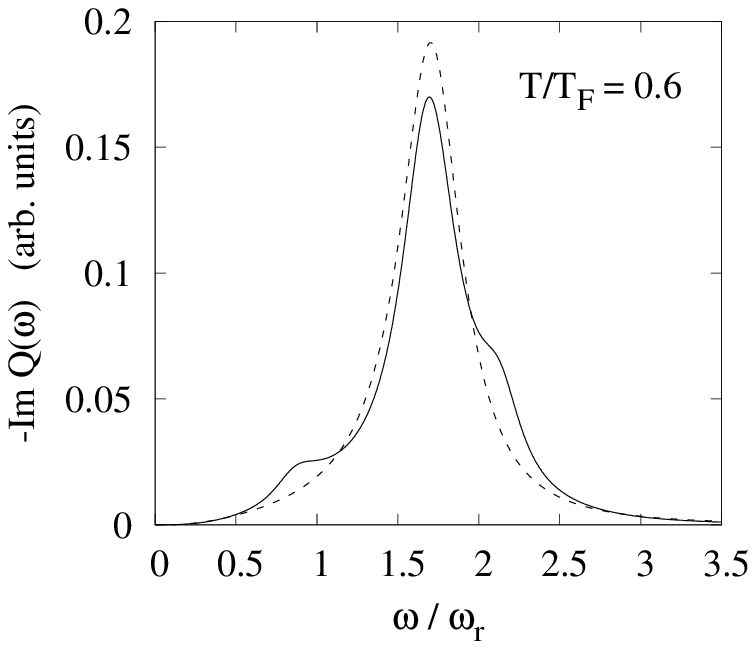} 
\includegraphics[width=5.9cm]{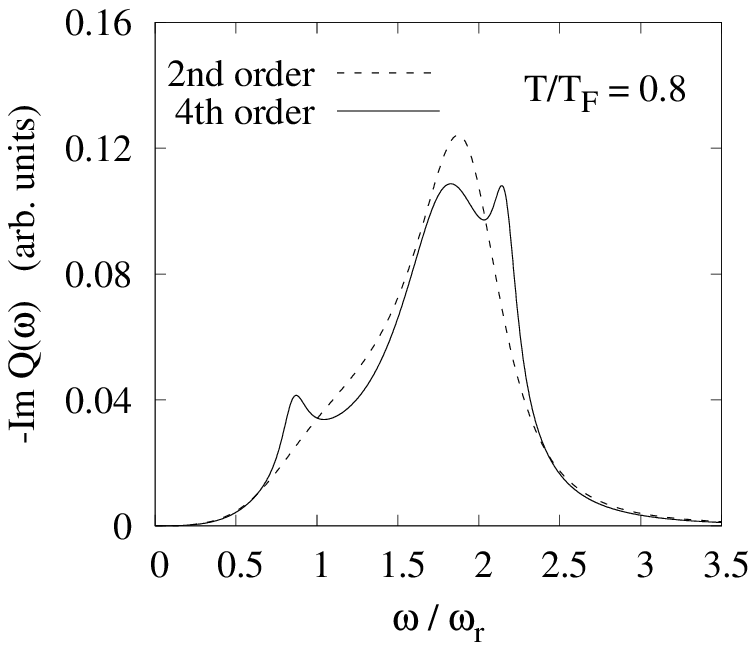}\\
\includegraphics[width=5.9cm]{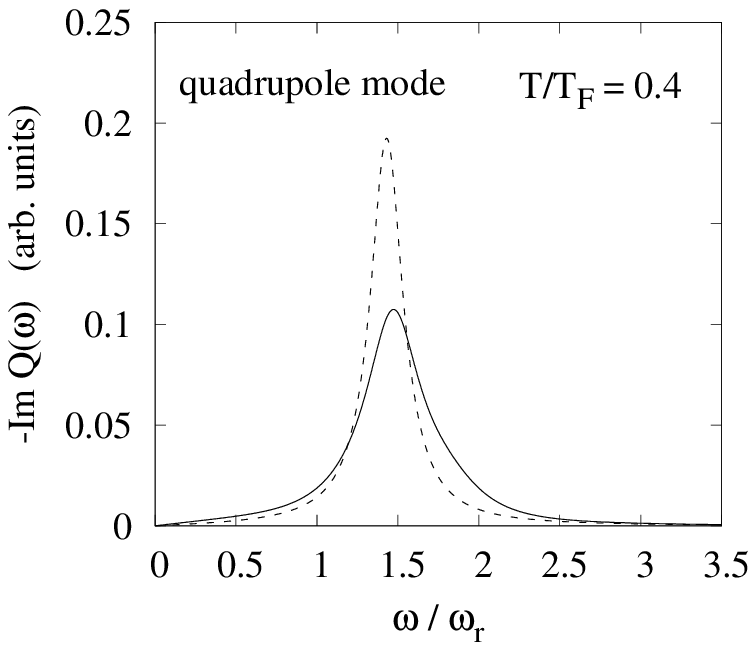}
\includegraphics[width=5.9cm]{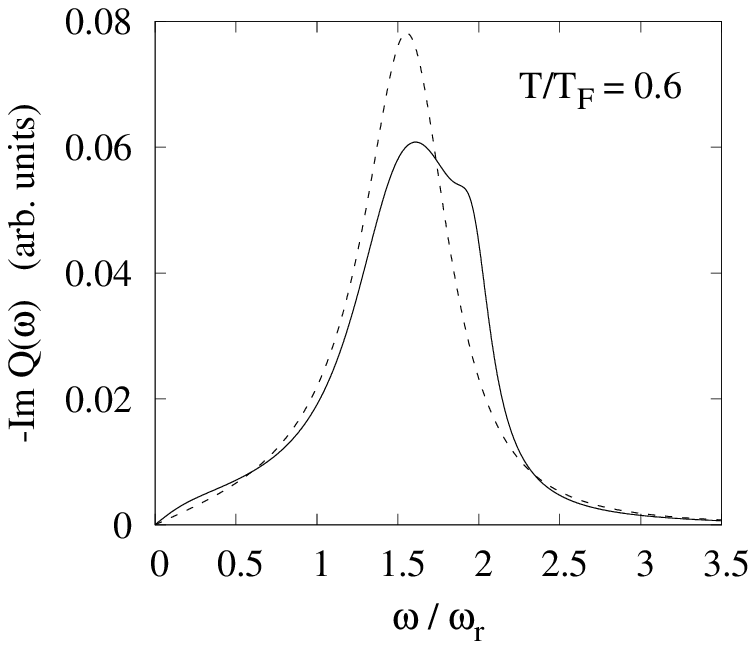} 
\includegraphics[width=5.9cm]{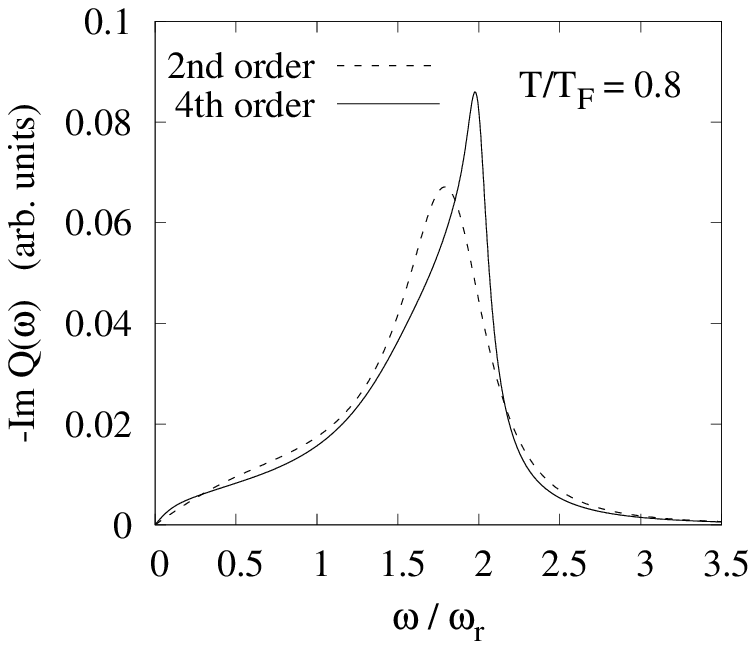}
\caption{\label{fig:Q} The imaginary part of the scissors (first row)
  and quadrupole (second row) response function as function of the
  frequency at temperatures $T/T_F = $ 0.4, 0.6, 0.8 (from left to
  right). The dashed lines represent the second-order results, the
  full ones the fourth-order results. The frequency is in
  units of the radial trap frequency $\omega_r=
  \sqrt{\omega_x\omega_y}$. The trap parameters are listed in \Tab{tab}.}
\end{figure*}

In \Fig{fig:Q}
we plot the results for $\Im Q(\omega)$ obtained at second and fourth
order for the scissors and quadrupole modes at unitarity. The
parameters of the trap and the number of $^6$Li atoms are chosen as in
\Ref{RiedlBruun}, so that a comparison with the experimental data is
possible, see \Tab{tab}.
\begin{table}
\caption{\label{tab}Trap parameters of the Innsbruck experiments. Both
  experiments were done with $600\,000$ atoms of $^6$Li in the unitary
  limit ($1/k_Fa = 0$) \cite{RiedlBruun}.}
\begin{ruledtabular}
\begin{tabular}{cccc}
mode & $\omega_x/2\pi$ (Hz) & $\omega_y/2\pi$ (Hz) &
  $\omega_z/2\pi$ (Hz)\\ 
\hline 
scissors   & 1600 & 700 & 30\\
quadrupole & 1800 & 1800 & 32
\end{tabular}
\end{ruledtabular}
\end{table}
In the upper panels of \Fig{fig:Q}, the scissors response is plotted
at various temperatures ($T/T_F= $ 0.4, 0.6, 0.8). Since mean-field
effects are not taken into account, the limiting frequencies for the
scissors mode in the hydrodynamic and collisionless regimes are
$\omega_{S,\hd}=\sqrt{\omega_x^2+\omega_y^2}$ and
$\omega_{S,\cl\pm}=\omega_x\pm\omega_y$, respectively
\cite{GueryOdelin}. (In the collisionless regime, two different modes
can be excited.) In the trap under consideration, these frequencies
are $\omega_{S,\hd}\simeq 1.65~\omega_r$, $\omega_{S,\cl-}\simeq
0.85~\omega_r$, and $\omega_{S,\cl+}\simeq 2.17~\omega_r$, where
$\omega_r = \sqrt{\omega_x\omega_y}$ is the average radial
frequency. Let us first analyse the second-order (dashed) curves. At
$T/T_F=0.4$, the response is peaked almost at $\omega_{S,hd}$: we are
in the hydrodynamic regime. As the temperature increases, the peak
becomes broader (strong damping) and gets shifted towards the higher
frequency $\omega_{S,\cl+}$. At second order, the lower mode at
$\omega_{S,\cl-}$ is not yet visible at $T/T_F = 0.8$ since it is
still too strongly damped. The fourth-order results (full lines)
deviate more and more from the lowest order ones as the temperature
increases. The most striking feature is that the shape itself of the
response function is modified by the inclusion of the higher-order
moments. We also observe that at fourth order the lower peak at
$\omega_{S,\cl-}$ is already clearly visible at $T/T_F = 0.8$.

In the second row of \Fig{fig:Q}, we plot the results for the
quadrupole mode. The limiting frequencies of this mode in the
hydrodynamic and collisionless limits are
$\omega_{Q,\hd}=\sqrt{2}\omega_r$ and $\omega_{Q,\cl}=2\omega_r$,
respectively (again without mean-field). The second order (dashed)
results show how the peak moves from the hydrodynamic to the
collisionless limit as the temperature increases. Consider now the
fourth-order (full) lines. At $T/T_F=0.4$ and $T/T_F=0.8$ the response
shows a clear peak, whose position is however displaced towards higher
frequencies, as compared to the second-order results.  At $T/T_F=0.6$,
the shape of the peak itself is deformed, but again its centroid is
moved towards higher frequencies. This is in qualitative agreement
with our finding in \Ref{Lepers} that the second-order moments method
overestimates the collisional effects, i.e., the second-order result
is always too close to the hydrodynamic limit.
\subsection{Frequencies and damping rates}
\begin{figure*}
\includegraphics[width=5.9cm]{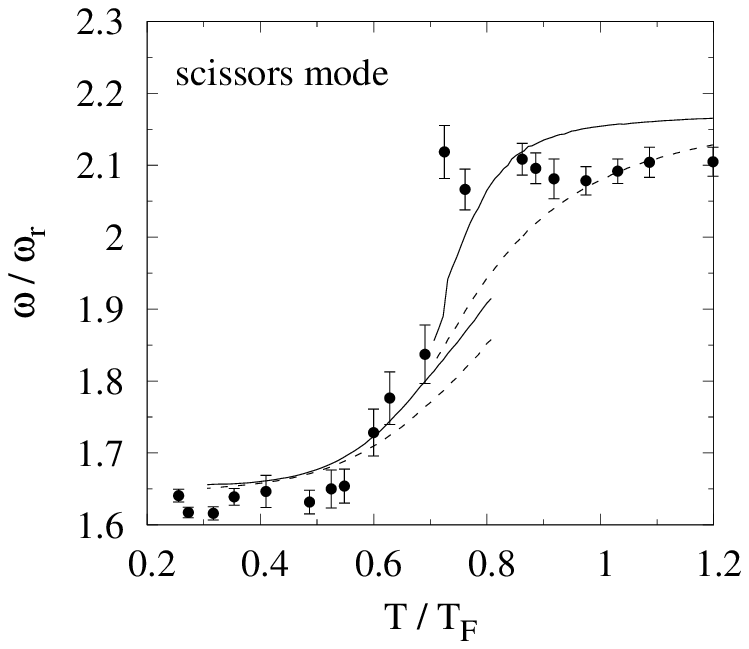} 
\includegraphics[width=5.9cm]{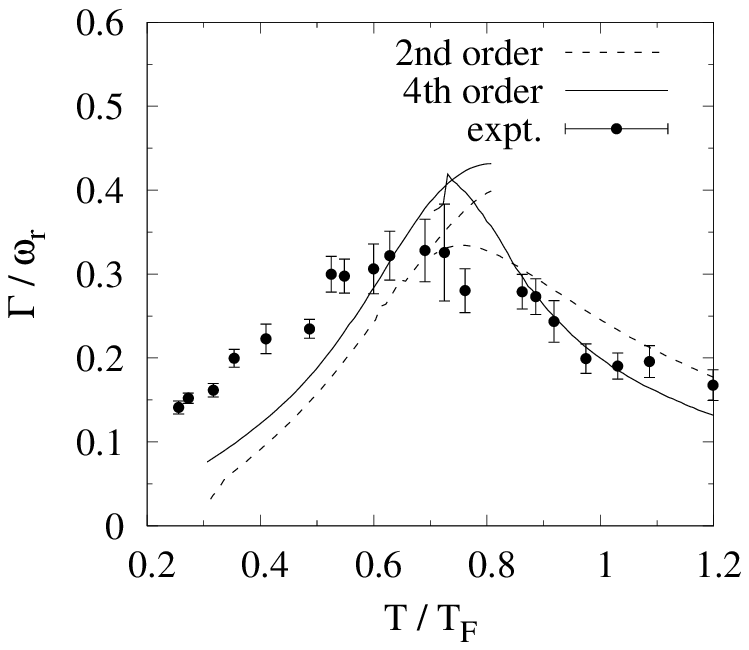} 
\includegraphics[width=5.9cm]{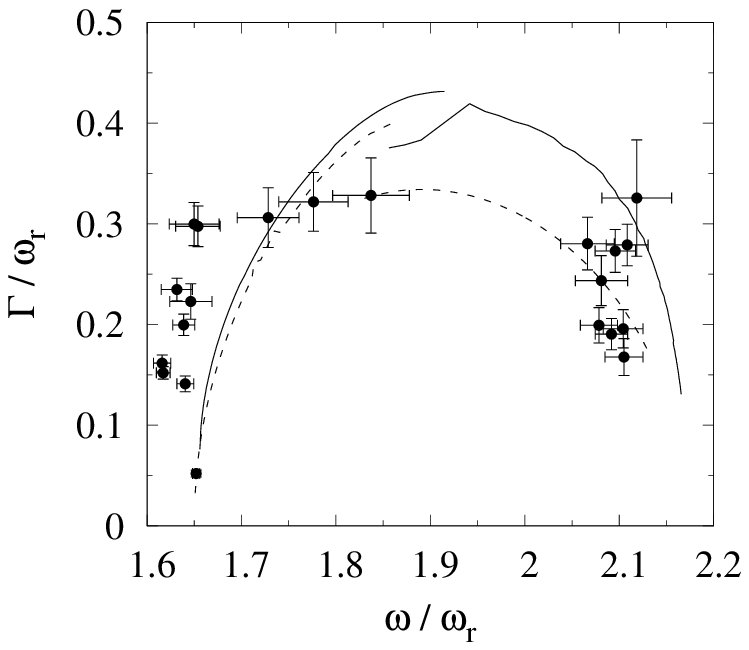}\\
\includegraphics[width=5.9cm]{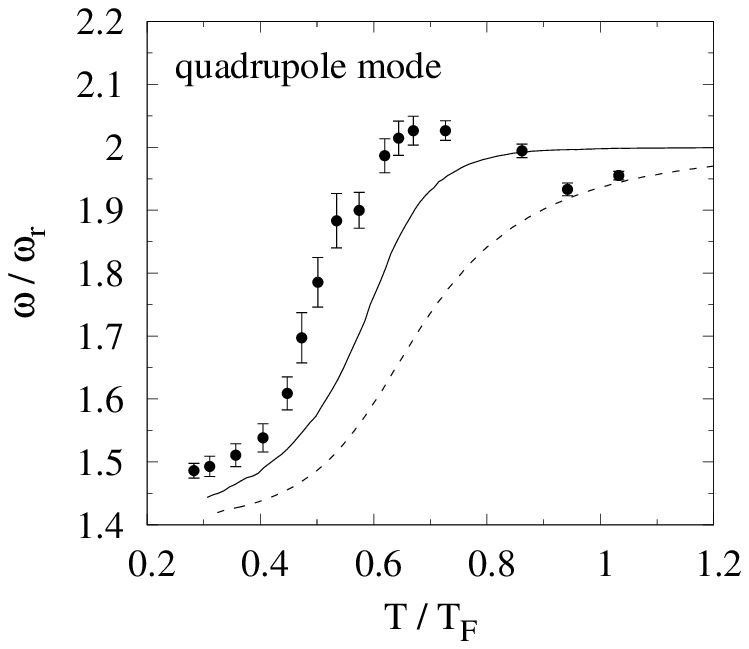}
\includegraphics[width=5.9cm]{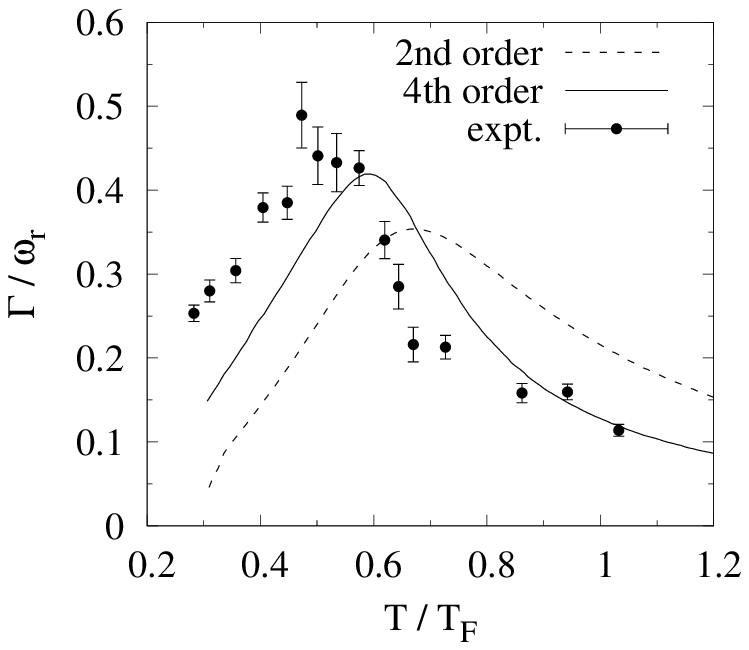}
\includegraphics[width=5.9cm]{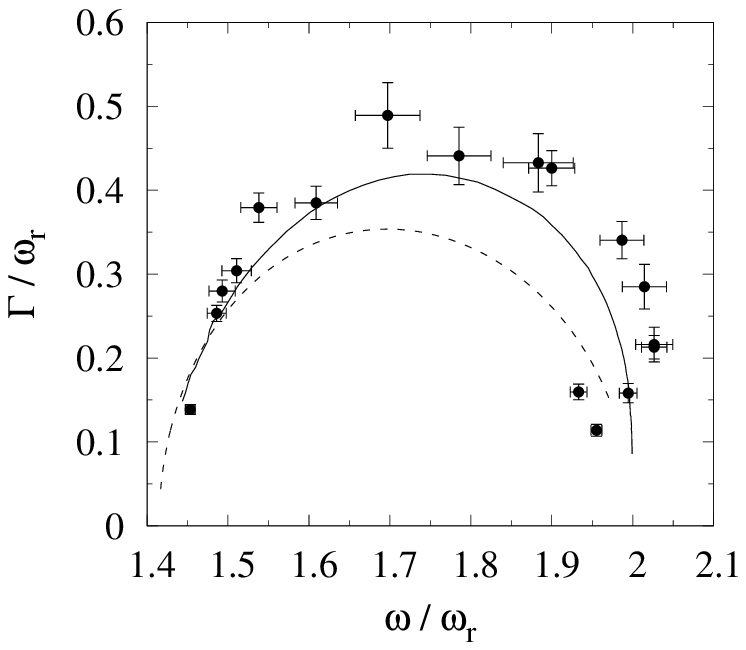}
\caption{\label{fig:omega} Frequency (left) and damping rate (middle) 
  of the scissors (first row) and radial quadrupole modes (second row)
  as functions of temperature as well as the representation damping
  vs. frequency (right) which is independent of possible
  uncertainties in the temperature measurement. The points with error
  bars are the experimental data from \Ref{RiedlBruun}, the dashed
  lines are the second-order results, and the solid lines are
  fourth-order results. The trap parameters are listed in \Tab{tab}.}
\end{figure*}

In order to make a quantitative comparison of our results with the
data, we extract from $Q(\omega)$ the frequency and damping of the
mode by fitting the response function as explained in \Sec{sec:fit}.
The results are shown in \Fig{fig:omega}.
Let us first look at the results for the scissors mode (first
row). Because of the two different fits at low and high temperatures
(using one or two damped cosine functions), there are two curves for
each method (second and fourth order moments). In the range of $T/T_F$
between 0.7 and 0.8 we plot both curves in order to show the
dependence on the fit. For $T/T_F > 0.8$, the gas is closer to the
collisionless regime where two modes are present, and we keep only the
fit with two damped cosine functions. For $T/T_F < 0.7$, we show only
the fit with a single damped cosine function. Let us now compare the
results obtained with the second-order (dashed lines) and fourth-order
(solid lines) moments methods. The most important difference is that
the transition from low frequency (hydrodynamic regime) to high
frequency (collisionless regime) is shifted to lower temperature by
the inclusion of fourth-order moments. This was to be expected since,
as we discussed above, the second-order moments method overestimates
the collisional effects. Therefore, for temperatures below $0.9 T_F$,
the fourth-order frequencies are in better agreement with the data
than the second-order ones. Only at high temperatures, it seems that
the second-order frequency, which approaches the limiting value
$\omega_{S,\cl+}$ much more slowly, is in better agreement with the
data. Concerning the damping, there is quite a big difference between
the second- and fourth-order results, but it is not really clear
whether the fourth-order represents an improvement or not.

The improvement due to the fourth-order moments is more clearly seen
in the results for the quadrupole mode (second row of
\Fig{fig:omega}). Again, if the fourth-order moments are included, the
transition from the hydrodynamic to the collisionless regime happens
at lower temperatures, which greatly improves the agreement of both
frequencies and damping rates with the data. But the difference
between second- and fourth-order calculations does not only concern
the temperature dependence. This can clearly be seen in the right
figure, showing the damping as function of frequency, so that the
temperature drops out. In this representation, the curve obtained with
the fourth-order moments almost passes through the error bars of the
data, which was by far not the case for the second-order results.

\section{Summary and conclusions}
\label{sec:summary}
In this work, we determined approximate solutions of the linearized
Boltzmann equation for collective modes of trapped Fermi gases by
using the method of phase-space moments. Here, we concentrated on the
radial quadrupole and scissors modes. Contrary to previous literature
\cite{BruunSmith07,RiedlBruun,Chiacchiera}, we did not only include the
lowest (second) order moments which are necessary to describe the
modes, but also the next (fourth) order. A comparison with a numerical
solution of the Boltzmann equation \cite{Lepers} showed that the
fourth order catches already the most important effects missed at
second order, e.g., the position dependence of the Fermi-surface
deformation. We therefore decided to apply this method to realistic
cases in order to be able to compare with experimental data.

We showed that, if one includes higher than second-order moments, the
shape of the response function does no longer resemble a single
Lorentzian. Therefore, if one wants to extract the frequency and
damping rate of a mode, the result depends on the ansatz for the fit
function which is used. Our determination of these quantities is
inspired by the procedure used by the experimentalists.

In the actual calculation of the moments of the collision term,
$A^\coll$, we used the in-medium cross section as defined in
\Ref{Chiacchiera}. In previous works \cite{RiedlBruun,Chiacchiera} it
was found that the in-medium enhancement of the cross section 
strongly deteriorates the agreement with the experimental
results. This conclusion was, however, based on calculations using
only second-order moments. Since the fourth-order contributions reduce
the effect of collisions, the effect of the enhanced cross section is
partially compensated. In fact, our new results, including the
in-medium cross section and fourth-order moments, are in reasonable
agreement with the data.

Another application of higher-order moments will be to quantify the
effects of the anharmonicity of the trap potential, including also the
mean field. Work in this direction is in progress. This might be
helpful, e.g., for understanding the behavior of the frequencies and
damping rates at high temperature (note that in \Ref{RiedlBruun} the
experimental frequencies have been roughly corrected for anharmonicity
effects by dividing them by the measured frequencies of the sloshing
mode). For a detailed comparison with the experiment, however, many
other effects should be accounted for, too. For instance, the measured
quantity is not the response to a $\delta$ pulse, but the relaxation
after the system was adiabatically deformed and then suddenly released
at $t=0$. This results, roughly speaking, in an additional factor
$1/\omega$ in the Fourier transform of the response of the system
which can have some effect on the fitted frequency and damping
rate. In addition, the observable measured in the experiment is not
simply proportional to $\langle x y\rangle$, but depends also on the
distribution in momentum space since the density profile in the $xy$
plane is measured after an expansion. Without any doubt, it would be
desirable to make a complete numerical simulation of the experiment,
including the expansion phase.
\section*{Acknowledgments}
S.C. is supported by FCT (Portugal) under the project SFRH/BPD/64405/2009.
\appendix
\section{Computation of $A_{ij}$}
In this appendix we give some details about the computation of the
matrix $A$ defined in \Eq{eq:Aij} and on its final form. 

As in our practical calculations, we will use trap units, i.e., all
quantities are made dimensionless by rescaling them by appropriate
combinations of the atom mass $m$, the average trap frequency
$\bar{\omega}=(\omega_x\omega_y\omega_z)^{1/3}$, the harmonic
oscillator length $l_\ho = 1/\sqrt{m\bar{\omega}}$ etc.
\subsection*{The matrix $M$}
In order to compute $M_{ij}$ defined in \Eq{eq:Mij}, it is convenient
to define six-dimensional hyperspherical coordinates. To do this, we
must first pass to isotropic spatial coordinates, and then to
dimensionless ones, so that the $\vek{r}$ and $\vek{p}$ components can
be treated together. We define
\begin{align}
x&=l_\ho (\bar{\omega}/\omega_x) X \cos\vartheta_1\nonumber\\
y&=l_\ho (\bar{\omega}/\omega_y) X \sin\vartheta_1 \cos\vartheta_2\nonumber\\
z&=l_\ho (\bar{\omega}/\omega_z) X \sin\vartheta_1 \sin\vartheta_2
  \cos\vartheta_3\nonumber\\
p_x&=(1/l_\ho) X \sin\vartheta_1 \sin\vartheta_2
  \sin\vartheta_3 \cos\vartheta_4\nonumber\\
p_y&= (1/l_\ho) X \sin\vartheta_1 \sin\vartheta_2
  \sin\vartheta_3 \sin\vartheta_4 \cos\varphi\nonumber\\
p_z&=(1/l_\ho) X \sin\vartheta_1 \sin\vartheta_2 \sin\vartheta_3
  \sin\vartheta_4 \sin\varphi\,.
\end{align}
The volume element becomes $d^3rd^3p=X^5 dX d\Omega_5$, and its angular
part is
\begin{equation}
d\Omega_5=\sin^4\vartheta_1\sin^3\vartheta_2\sin^2\vartheta_3\sin\vartheta_4
d\vartheta_1 d\vartheta_2 d\vartheta_3 d\vartheta_4 d\varphi\,.
\end{equation}
The integration range is $[0,\infty[$ for $X$, $[0,2\pi]$ for
$\varphi$ and $[0,\pi]$ for the $\vartheta_i$. In these
coordinates, the equilibrium distribution function reduces to
\begin{equation}
f_{eq}(X)=\frac{1}{e^{\beta(\bar{\omega}X^2/2-\mu)}+1}~,
\end{equation}
and one obtains the useful relation
\begin{equation}
  \frac{d f_\eq(X)}{d X}=-f_\eq (1-f_\eq)\beta\bar{\omega}X~.
\end{equation}
Using the latter, one can check that 
\begin{equation}
\int d\Omega_5\int_0^\infty\frac{dX}{(2\pi)^3} X^5f_\eq(1-f_\eq)X^n=
\frac{n+4}{\beta\bar{\omega}}\frac{N}{2}\langle X^{n-2}\rangle_\eq~.
\end{equation}
Then, it can easily be seen that the elements of $M$ are proportional
to $\langle X^n\rangle_\eq$, $n=2,4,6$. We choose to express them in
terms of $\langle x^n\rangle_\eq $: in trap units, the factors of
proportionality contain the factor $N/\beta$, rational numbers and
ratios of powers of the trap frequencies.
\subsection*{The matrix $A^\trans$}
Notice that in the case of a harmonic trap, the set $\{\phi_i\}$ is
closed with respect to the operators in the left-hand side of \Eq{eq:BoltzLin},
therefore one can find a matrix $B$ of coefficients such that
\begin{equation}
\{\phi_j,\frac{p^2}{2m}+V\}=\sum_{k=1}^n \phi_k B_{kj}\,.
\end{equation}
Then, it is clear that the matrix $A^\trans$ defined in
\Eq{eq:Atransij} can be written as a matrix product
\begin{equation}\label{eq:Acomp}
A^\trans = M B\,,
\end{equation}
where $M$ denotes the matrix calculated in the preceding subsection.

The computation of the matrix $B$ is straight-forward. In trap units,
its elements are simply given by powers of the trap frequencies
multiplied by integer numbers.
\subsection*{The matrix $A^\coll$}
To compute the matrix elements $A^\coll_{ij}$, that by definition are
\begin{multline}\label{eq:Acoll}
A^\coll_{ij}=\int\frac{d^3rd^3p}{(2\pi)^3} \phi_i(\rv,\pv)
\int \frac{d^3 p_1}{(2\pi)^3}\int d\Omega
\frac{d\sigma}{d\Omega} \frac{|\vek{p}-\vek{p}_1|}{m}\\
\times f_\eq f_{\eq\,1}
(1-f_\eq^\prime)(1-f_{\eq\,1}^\prime)
\Delta_\coll[\phi_j]\,,
\end{multline}
we follow the method outlined in \Refs{Vichi,Chiacchiera}.  In the last
equation we have used the compact notation $\Delta_\coll[\phi]=
\phi(\vek{r},\vek{p})+\phi(\vek{r},\vek{p}_1)
-\phi(\vek{r},\vek{p}^\prime)-\phi(\vek{r},\vek{p}_1^\prime)$.  To
reduce the number of integrals in \Eq{eq:Acoll}, we define the
variables $\vek{k} = \vek{p}+\vek{p}_1$,
$\vek{q} = (\vek{p}-\vek{p}_1)/2$ and
$\vek{q}^\prime=(\vek{p}^\prime-\vek{p}_1^\prime)/2$ (remember
that energy and momentum conservation imply
$|\vek{q}|=|\vek{q}^\prime|$). In these variables, one can write
\begin{multline}
f_\eq f_{\eq\,1}(1-f_\eq^\prime)(1-f_{\eq\,1}^\prime) = \\
\frac{1}{4}\,\frac{1}{\cosh \beta(E-\mu)+
  \cosh \beta\vek{k}\cdot\vek{q}/2m}\\
\times \frac{1}{\cosh \beta(E-\mu)+
  \cosh \beta\vek{k}\cdot\vek{q}^\prime/2m}\,,
\end{multline}
with $E = k^2/8m+q^2/2m+V$.
The factor $\phi_i\Delta[\phi_j]$ has to be computed and rewritten, as
the rest of the integrand, in these variables, too. Then, we define a
rotation that brings $\vek{k}$ (identified by the angles
$\theta,\varphi$) to be parallel to the $z$-axis in momentum space. We
define $R$ the matrix associated to such a rotation and apply it to
all momenta: the old coordinates are related to the new ones by
$(p_x,p_y,p_z)=R^{-1}(p_a,p_b,p_c)$, and in particular
$(k_a,k_b,k_c)=(0,0,k)$. Now the integration over $\theta,\varphi$ can
be performed analytically, since all the dependence upon these
variables is in the numerator of the integrand. We have thus reduced
the number of integrals from eleven to nine. Next one defines
spherical coordinates for $\vek{q}$ and $\vek{q}^\prime$: their zenith
and azimuth angles are $\theta_c,\varphi_c$ and
$\theta_c^\prime,\varphi_c^\prime$ respectively. Since the dependence
upon $\varphi,\varphi_c^\prime$ is only in the numerator, we can
easily integrate over these variables, reducing the integral to a
seven-dimensional one. Finally, the definition of scaled spatial
coordinates $\tilde{r}_i\equiv \frac{\omega_i}{\bar{\omega}}r_i$
renders the trap potential, and therefore the integrand, spherically
symmetric in the spatial coordinates: the integral is reduced to a
five-dimensional one. As a result, the elements of $A^\coll$ are
proportional, through rational numbers and ratios of powers of trap
frequencies, to terms of the same type of the inverse relaxation time
$1/\tau$ defined in \Ref{Chiacchiera}. More precisely, now there are
twelve different terms of this type which are of the form
\begin{multline}
J_i = \frac{1}{20\pi^2m} \int_0^\infty d\tilde{r}\, \tilde{r}^2 dk\, k^2 dq\,
   q^7 \frac{d\sigma}{d\Omega} \int_{-1}^1 d\gamma d\gamma^\prime F_i\\
\times\frac{1}{\cosh\beta(E-\mu) +\cosh\beta k q \gamma/2m}\\
\times\frac{1}{\cosh\beta(E-\mu) +\cosh\beta k q \gamma^\prime/2m}~,
\end{multline}
$i=1,\dots,12$, where $E=k^2/8m+q^2/2m+m\bar\omega^2\tilde{r}^2/2$,
$\gamma=\cos\theta_c$ and
$\gamma^\prime=\cos\theta_c^\prime$.  The factors $F_i$ are
polynomials of $\tilde{r}^2$, $k^2$, $q^2$, $\gamma^2$, and
${\gamma^\prime}^2$.
In particular, $F_1 = 1+2\gamma^2-3\gamma^2\gamma^{\prime\, 2}$, such
that $J_1$ is identical to $I_S$ given in Eq. (B4) of
\Ref{Chiacchiera}, and is in fact the only non-zero term of
$A^\coll$ at second order. The coefficients $J_i$ are obtained
numerically via a Monte Carlo integration and used to build
$A^\coll$.
\section{Calculation of the response function}
In this appendix we describe how the Fourier spectrum of a generic
observable $\langle \chi\rangle$ after a generic perturbation $\hat{V}$
can be obtained.

First, one has to calculate the vector $a_i$ defined in \Eq{eq:ai}
[which is simple in the case $\hat{V}=xy$, cf. \Eq{eq:aiexplicit}].
Then, one has to express the expectation value of $\chi$ in terms of the
coefficients $c_i$. Supposing that $\langle \chi\rangle_\eq=0$, the
expectation value must be proportional to the $c_i$ and one can thus
write
\begin{equation}
\langle \chi\rangle = \sum_{i=1}^n b_i c_i = b^T c\,,
\end{equation}
where we changed to vector notation in the second equality, $b$ and
$c$ being vectors with components $b_i$ and $c_i$, respectively.

Inverting \Eq{eq:Ac=a}, one obtains
\begin{align}
\langle \chi\rangle (\omega) & = b^T (-i\omega M + A^\trans +
  A^\coll)^{-1} a\nonumber\\
  & = b^T [-i\omega\openone + M^{-1}(A^\trans+A^\coll)]^{-1} M^{-1}a\,.
\end{align}
Notice that $M$, $A^\trans$, and $A^\coll$ are real matrices which are
independent of $\omega$. Now we perform the diagonalization
\begin{equation}
M^{-1}(A^\trans+A^\coll)=PDP^{-1}\,,
\end{equation}
with $D=\mathrm{diag}(\lambda_1,\dots,\lambda_n)$. Since the original
matrix is real, its eigenvalues $\lambda_k$ are either real or
they appear as complex conjugate pairs. If we identify the real
and imaginary parts of the eigenvalues as $\lambda_k =
\Gamma_k+i\omega_k$, we obtain
\begin{align}
\langle \chi\rangle (\omega) & = b^T P(-i\omega
\openone+D)^{-1} P^{-1}M^{-1}a\nonumber\\ 
  & = i \sum_{k=1}^n \frac{(b^T P)_k
  (P^{-1}M^{-1}a)_k}{\omega-\omega_k+i\Gamma_k}\,.
\end{align}
This reduces to \Eq{responsefunction} in the special case $\hat{V} = \chi
= xy$.


\end{document}